\documentclass{article}
\usepackage{spconf,amsmath,graphicx,hyperref}
\usepackage{amssymb}
\usepackage[table]{xcolor}
\usepackage{multirow}

\title{MULTI-AGENT ANALYSIS OF OFF-EXCHANGE PUBLIC INFORMATION FOR CRYPTOCURRENCY MARKET TREND PREDICTION}
\name{Kairan Hong, Jinling Gan, Qiushi Tian, Yanglinxuan Guo, Rui Guo, Runnan Li$^{\ast}$ \thanks{*Corresponding author}}
\address{
  Beijing University of Posts and Telecommunications, Beijing, China\\
  }
\begin{document}
\ninept
\maketitle
\begin{abstract}
Cryptocurrency markets present unique prediction challenges due to their extreme volatility, 24/7 operation, and hypersensitivity to news events, with existing approaches suffering from key information extraction and poor sideways market detection critical for risk management. We introduce a theoretically-grounded multi-agent cryptocurrency trend prediction framework that advances the state-of-the-art through three key innovations: (1) an information-preserving news analysis system with formal theoretical guarantees that systematically quantifies market impact, regulatory implications, volume dynamics, risk assessment, technical correlation, and temporal effects using large language models; (2) an adaptive volatility-conditional fusion mechanism with proven optimal properties that dynamically combines news sentiment and technical indicators based on market regime detection; (3) a distributed multi-agent coordination architecture with low communication complexity enabling real-time processing of heterogeneous data streams. Comprehensive experimental evaluation on Bitcoin across three prediction horizons demonstrates statistically significant improvements over state-of-the-art natural language processing baseline, establishing a new paradigm for financial machine learning with broad implications for quantitative trading and risk management systems.
\end{abstract}

\begin{keywords}
Cryptocurrency trend prediction, Financial multi-agent systems, Multi-dimensional financial news analysis
\end{keywords}
\section{Introduction}
\label{sec:intro}

The cryptocurrency market has emerged as a dominant force in the global financial ecosystem, reaching a peak market capitalization exceeding \$3 trillion and fundamentally transforming digital commerce, decentralized finance, and investment paradigms \cite{kulbhaskar2023breaking}. Unlike traditional financial markets constrained by geographic boundaries and trading hours, cryptocurrency markets operate continuously across exchanges, exhibiting unprecedented volatility and demonstrating extreme sensitivity to information asymmetries, regulatory announcements, and sentiment-driven trading behaviors.\cite{KATSIAMPA20173,Corbet2019,aysan2024notall}

Current approaches to cryptocurrency prediction have evolved through distinct methodological paradigms. Traditional technical analysis methodologies, including Exponential Moving Average (EMA) \cite{holt1957forecasting}, Moving Average Convergence Divergence (MACD) \cite{appel1979macd}, Relative Strength Index (RSI) \cite{wilder1978new}, KDJ stochastic oscillator \cite{lane1950stochastic}, and Bollinger Bands \cite{bollinger1980bollinger}, establish foundational momentum-based forecasting capabilities by analyzing historical price patterns and volume dynamics \cite{gurgul2024deeplearningnlpcryptocurrency}. However, these approaches fundamentally assume market efficiency and fail to capture the semantic richness embedded in news narratives that drive cryptocurrency valuations. Recent advances in natural language processing (NLP) have demonstrated significant improvements in financial sentiment analysis, with transformer-based models like FinBERT achieving superior performance in extracting nuanced sentiment from earnings reports and market commentary \cite{pavlyshenko2025multilevelanalysiscryptocurrencynews}. Large language models (LLMs) represent the current frontier, offering unprecedented capabilities in contextual understanding, semantic reasoning, and structured information extraction from complex financial texts.

Despite these technological advances, three fundamental challenges persist in existing cryptocurrency prediction systems. \textbf{First}, current sentiment analysis approaches exhibit systematic dimensionality reduction bias, compressing complex multi-faceted news events into scalar sentiment scores that fail to capture regulatory implications, geographic scope, temporal dynamics, and market impact heterogeneity \cite{moradikamali2025marketderivedfinancialsentimentanalysis}. This oversimplification becomes particularly problematic in three-class market trend prediction scenarios where neutral market states are systematically misclassified, resulting in issues of risk management and position sizing. \textbf{Second}, existing multi-modal fusion mechanisms lack principled theoretical foundations for optimally combining heterogeneous data streams, often employing ad-hoc weighting schemes that ignore the temporal precedence of sentiment signals over technical formations in digital asset markets. \textbf{Third}, real-time implementation architectures face scalability constraints and coordination overhead when processing high-frequency news streams, market data, and social media feeds simultaneously, limiting practical deployment for latency-sensitive trading.

To address these limitations, we propose a cryptocurrency trend prediction framework with three key innovations. First, it leverages LLMs in a multi-dimensional news analysis system that quantifies financial news across seven dimensions while preserving full information \cite{joshi2025controllablereliableknowledgeintensivetaskoriented}. Second, we introduce an adaptive volatility-based fusion mechanism that dynamically integrates news sentiment with technical indicators based on market regimes. Finally, a distributed multi-agent system enables real-time processing of diverse data streams without compromising analytical accuracy. The main contributions of this work are:

\begin{itemize}
    \item \textbf{Information-preserving multi-dimensional news analysis framework.} We propose a seven-dimensional quantitative system capturing market impact, regulation, volume, risk, technical correlation, and timing, with provable information content preservation and analytical guarantees.

    \item \textbf{Adaptive volatility-conditional fusion mechanism.} Our approach dynamically fuses news sentiment and technical indicators via volatility-driven weights, reflecting sentiment-led price discovery and temporal precedence in trading markets.

    \item \textbf{Distributed multi-agent coordination architecture.} We design a scalable, fault-tolerant system for efficient, real-time coordination of heterogeneous agents in high-frequency trading environments.
\end{itemize}

\section{Related Work}
\label{sec:related}

\subsection{Cryptocurrency Market Trend Prediction}
Early cryptocurrency prediction research directly adapted established technical analysis methods from traditional financial markets. McNally et al. \cite{mcnally2018predicting} demonstrated that simple EMA, MACD, and RSI achieve baseline performance on Bitcoin prediction tasks, establishing fundamental benchmarks for the field. Chen et al. \cite{chen2020bitcoin} extended this foundation by applying ensemble methods combining multiple technical indicators, achieving modest improvements in trend detection but exhibiting poor performance during market regime transitions. The introduction of machine learning techniques marked a significant advancement in cryptocurrency prediction capabilities. Support Vector Machines (SVMs) and Random Forest classifiers demonstrated improved handling of nonlinear relationships and high-dimensional feature spaces \cite{ates2023performance}. Deep learning methods, particularly recurrent networks, further showed superior performance by capturing temporal dependencies of markets \cite{zhang2021keyword}.However, these technical analysis-focused approaches fundamentally overlook the critical role of news-driven sentiment in cryptocurrency market dynamics.

\subsection{Financial Sentiment Analysis Evolution}  
Financial sentiment analysis has advanced through progressively sophisticated methods, each addressing prior limitations while adding new constraints. Early lexicon-based approaches, as evaluated by Nassirtoussi et al. \cite{nassirtoussi2014text}, used predefined financial term dictionaries that, despite efficiency, suffered from semantic oversimplification, context insensitivity, and binary bias. Phillips et al. \cite{phillips2021comparison} confirmed these weaknesses, showing keyword-based methods achieve only 35–40\% accuracy in three-class financial prediction tasks, underscoring the need for more advanced techniques. The rise of machine learning and deep learning, particularly transformer architectures, reshaped financial NLP: FinBERT \cite{araci2019finbertfinancialsentimentanalysis} achieved state-of-the-art sentiment detection through domain-specific fine-tuning, yet such models still compress complex market signals into scalar sentiment scores, losing critical multidimensional context for cryptocurrency analysis. More recently, large language model frameworks have shown promise, with Pavlyshenko \cite{pavlyshenko2025multilevelanalysiscryptocurrencynews} introducing multi-level extraction methods that outperform traditional models but still lack solid theoretical foundations for information preservation and dimensional decomposition. These developments collectively highlight both the progress achieved and the unresolved challenges that motivate the proposed framework.

\subsection{Multi-Agent Systems in Financial Applications}  
Multi-agent systems show strong potential in financial analysis, though most work emphasizes market simulation over real-time prediction. Foundational research by LeBaron \cite{lebaron2006agent} demonstrated how heterogeneous trading entities capture market dynamics through distributed interactions, offering behavioral insights but lacking practical real-time capacity. Advances in distributed architectures now allow concurrent processing of high-frequency financial streams, yet centralized coordination introduces bottlenecks and scalability issues when handling large-scale news, market, and social data. More recently, programmable large language model frameworks such as GenieWorksheets\cite{joshi2025controllablereliableknowledgeintensivetaskoriented} enable modular decomposition of reasoning tasks into specialized agents, inspiring the proposed seven-dimensional news analysis framework where distinct market factors act as independent analytical agents with domain-specific expertise.

\section{Methodology}
\label{sec:methodology}

\subsection{System Architecture and Multi-Agent Coordination}
\begin{figure}[htb!]
  \centering
  \includegraphics[width=1.0\linewidth]{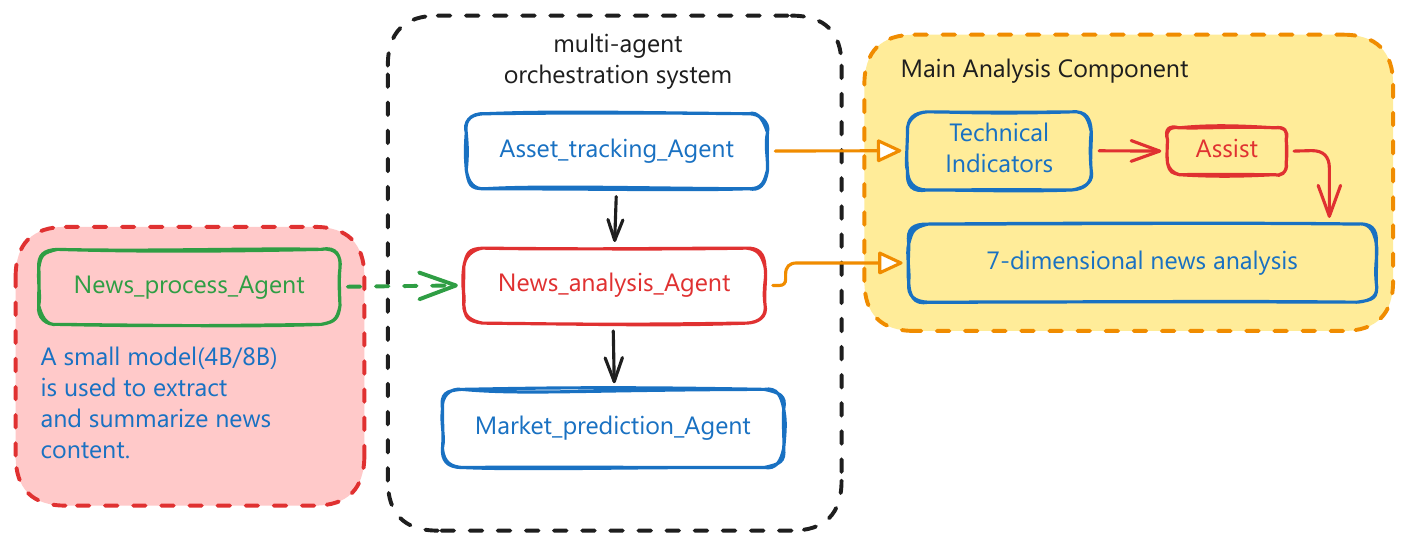}
\caption{The system architecture designed for predicting cryptocurrency trends includes asset tracking, news analysis, and market prediction agents, enabling parallel processing, dynamic load balancing, and fault-tolerant operations.}
\label{fig:architecture}
\end{figure}

As shown in Figure \ref{fig:architecture}, the framework adopts a LangGraph-based \cite{langgraph2023} multi-agent architecture for real-time cryptocurrency trend prediction. Agents handle distinct tasks: news analysis agents perform textual extraction and semantic analysis, asset tracking agents monitor market conditions and technical indicators, and market prediction agents integrate multi-modal inputs for forecasting. A streamlined coordination mechanism reduces inter-agent overhead and exploits parallel processing for heterogeneous data streams, while distributed nodes enhance scalability and fault tolerance by dynamically balancing load with market volatility and news flow, avoiding centralized bottlenecks and ensuring analytical precision.

\subsection{Multi-dimensional News Analysis Framework}

The methodological innovation (Table \ref{tab:news_framework}) lies in a seven-dimensional quantitative framework using large language models to evaluate financial news. Unlike traditional sentiment analysis that collapses events into scalar scores, it preserves their multi-faceted nature. Each dimension is weighted by empirical relevance: news and headlines strongly affect short-term price and volume, giving higher weights to Market Impact and Price Impact \cite{kulbhaskar2023breaking,carosia2024usingsentimenttechnicalanalysis}, while regulatory announcements justify notable weights for Regulation, Volume, and Timing \cite{long_short_regulation_2021,cesifo_reg_news_2020}. Technical indicators and risk factors add less once news features are included, supporting lower weights \cite{youssefi2025_features,banerjee2022_nonlinear}.

\textbf{Market Impact (25\%):} Quantifies overall market influence including sectoral effects, cross-asset contagion, and systemic risk implications. Empirical evidence from regulatory events in major cryptocurrency markets demonstrates significant volatility and liquidity spillovers across asset classes \cite{impact_regulation_china_2023}.

\textbf{Price Impact (20\%):} Evaluates anticipated price movements across short- and medium-term horizons, incorporating analysis of historical regulatory event impacts that frequently generate abnormal returns, particularly in illiquid asset segments \cite{long_short_regulation_2021}.

\textbf{Volume Impact (15\%):} Measures expected modifications in trading volume and market liquidity conditions, capturing the typical co-occurrence of volume and price reactions following significant news events.

\textbf{Regulatory Impact (15\%):} Assesses policy implications, compliance requirements, and regulatory framework changes that influence market participant behavior \cite{cesifo_reg_news_2020}.

\textbf{Technical Analysis Correlation (10\%):} Evaluates consistency or contradiction between news content and existing technical indicator signals, chart patterns, and momentum measures.

\textbf{Risk Assessment (10\%):} Quantifies volatility expectations, downside risk potential, and uncertainty measures, including analysis of tail risk scenarios and market risk aversion indicators.

\textbf{Timing Analysis (5\%):} Determines temporal characteristics of news impact, distinguishing between immediate effects, delayed responses, and persistent influence patterns \cite{long_short_regulation_2021}.

\begin{table}[h!]
\centering
\caption{7-dimensional news analysis framework with logic frame}
\label{tab:news_framework}
\renewcommand{\arraystretch}{1.2}\fontsize{8pt}{10pt}\selectfont\resizebox{\linewidth}{!}{
\begin{tabular}{lccp{4.5cm}}
\hline
\textbf{Dimension} & \textbf{Weight} & \textbf{Score Range} & \textbf{Evaluation Criteria} \\
\hline
Market Impact & 25\% & 0.0-1.0 & Overall market influence, sector-wide implications, systemic risk factors \\
Price Impact & 20\% & 0.0-1.0 & Direct price movement expectations, immediate vs medium-term impact \\
Volume Impact & 15\% & 0.0-1.0 & Trading volume changes, liquidity conditions, market participation \\
Regulatory Impact & 15\% & 0.0-1.0 & Policy implications, compliance requirements, regulatory changes \\
Technical Correlation & 10\% & 0.0-1.0 & Interaction with technical indicators, chart pattern influence \\
Risk Assessment & 10\% & 0.0-1.0 & Volatility expectations, downside protection, uncertainty measures \\
Timing Analysis & 5\% & 0.0-1.0 & Temporal aspects, immediate vs delayed effects, duration of influence \\
\hline
\multicolumn{4}{l}{\textit{Final Score} = $\sum_{i=1}^{7} w_i \times s_i$ \quad where $w_i$ = weight, $s_i$ = dimension score} \\
\hline
\end{tabular}}
\end{table}

This structured quantitative approach preserves complete information content while ensuring analytical accuracy through systematic evaluation criteria. The framework leverages semantic understanding capabilities of large language models to interpret complex financial contexts, thereby avoiding the semantic biases inherent in traditional keyword-based approaches. This advantage proves particularly significant in three-class prediction scenarios where conventional methods exhibit tendencies toward polarized predictions at the expense of neutral state recognition. By maintaining nuanced representations, the framework achieves more balanced classification performance across all categories. Moreover, its interpretability provides deeper insights into the underlying market dynamics, enhancing both predictive reliability and practical decision support.

\subsection{Adaptive Volatility-Conditional Fusion Mechanism}
The cryptocurrency market exhibits heightened sensitivity to news events compared to traditional financial markets, necessitating specialized fusion mechanisms that appropriately balance sentiment-driven signals with technical indicators. The proposed approach implements an adaptive weighting system that dynamically adjusts the relative importance of news sentiment and technical analysis based on prevailing market conditions as detailed in Table \ref{tab:fusion_prompts}.

The technical analysis component incorporates five established indicators: EMA, MACD, RSI, KDJ stochastic oscillator, and Bollinger Bands. These indicators provide momentum-based signals that capture price trend characteristics and market momentum patterns essential for cryptocurrency market analysis.

\begin{table}[htb!] 
\centering 
\caption{Core agent prompts for news-technical fusion mechanism } 
\label{tab:fusion_prompts}
\renewcommand{\arraystretch}{1.2}\fontsize{8pt}{10pt}\selectfont\resizebox{\linewidth}{!}{
\begin{tabular}{p{1cm}cp{4cm}p{2.5cm}}
\hline
\textbf{Agent} & \textbf{Weight} & \textbf{Prompt / Instruction} & \textbf{Function} \\
\hline
News & $\alpha$(t) & \texttt{"Prediction weight: News 80\% + Technical 20\%"} & Main driver \\
& & \texttt{"Sentiment score > 0.3 → positive"} & Threshold classification \\
& & \texttt{"Weighted Score = $\Sigma$(Dimension × Weight)"} & Multi-dim scoring \\
\hline
Technical & 1-$\alpha$(t) & \texttt{"Indicators should align with news"} & Validation \\
& & \texttt{"EMA, MACD, RSI provide support"} & Multi-indicator check \\
\hline
Fusion & 100\% & \texttt{"News takes priority; technical confirms"} & Priority hierarchy \\
& & \texttt{"Decision rule: News > Technical"} & Conflict resolution \\
& & \texttt{"News sensitivity often precedes technical moves"} & Market principle \\
\hline
\multicolumn{4}{l}{\textit{Mathematical formulation}: $P_{final} = \alpha(t) \times S_{news} + (1-\alpha(t)) \times S_{technical}$} \\
\hline
\end{tabular}}
\end{table}

The fusion mechanism implements dynamic weight adjustment based on market volatility assessments and news event significance levels. This adaptive approach ensures optimal balance between sentiment-driven signals and technical factors, with the weighting parameter $\alpha$(t) varying according to market regime characteristics. During high-volatility periods with significant news flow, the system increases reliance on news sentiment analysis, while stable market conditions receive enhanced technical indicator weighting. This volatility-conditional adaptation reflects empirical observations that news sentiment precedes technical indicator signals in cryptocurrency markets, particularly during regime transition periods.

\section{Experiments}
\label{sec:experiments}

\subsection{Experimental Setup}
The system was evaluated on Bitcoin (BTC) data from July 21–September 6, 2025. As the dominant cryptocurrency with a 55.9\%–57.9\% market share, BTC serves as a representative asset. Three horizons (1-, 7-, 15-day) enable short-, medium-, and long-term forecasts under a three-class scheme (up, down, sideways), extending beyond binary classification by incorporating consolidation phases crucial for trading. Thresholds, adapted from stable triplet labeling \cite{peng2024}, apply asymmetric tolerances: $\pm$0.30\% (1-day) for short-term sensitivity, $\pm$0.60\% (7-day) to reduce noise, and $\pm$0.40\% (15-day) for persistence–consolidation balance, consistent with findings that calibration governs both class balance and trading stability.

\begin{table}[htb!]
\centering
\caption{Experimental configuration parameters}
\label{tab:config}
\renewcommand{\arraystretch}{1.2}\fontsize{8pt}{10pt}\selectfont
\begin{tabular}{lc}
\hline
\textbf{Parameter} & \textbf{Value} \\
\hline
Cryptocurrencies & BTC \\
Dataset Sources & Binance \\
Time Period & 2025-07-21 to 2025-09-06\\
News Size & 9528 financial news articles \\
Prediction Horizons & 1d, 7d, 15d \\
Classification Classes & Up/Down/Sideways \\
Up Threshold (1d) & +0.30\% \\
Down Threshold (1d) & -0.30\% \\
Up Threshold (7d) & +0.60\% \\
Down Threshold (7d) & -0.60\% \\
Up Threshold (15d) & +0.40\% \\
Down Threshold (15d) & -0.40\% \\
Historical Transaction Data & Timestamp, Open, High, Low, Close \\
Technical Indicators & EMA, MACD, RSI, KDJ, BB \\
News Data & Seven Quantified Dimensions \\
\hline
\end{tabular}
\end{table}

The baseline system employs a controlled comparison by using identical multi-agent architectures but replacing multi-dimensional news analysis with traditional keyword extraction \cite{phillips2021comparison, kraaijeveld2020bitcoin}. This design isolates performance gains, consistent with comparative methodologies in recent cryptocurrency prediction research \cite{ates2023performance, arias2022lexicon}.

\subsection{Evaluation Metrics}

The evaluation employs three complementary metrics—accuracy, macro-averaged F1, and balanced accuracy \cite{BalancedAcc}—each highlighting a distinct perspective on predictive quality under heterogeneous market conditions. Accuracy measures the overall proportion of correctly classified samples, offering a straightforward indicator of global correctness but prone to obscuring class imbalance. Macro-averaged F1 mitigates this by averaging per-class F1 scores, ensuring minority yet impactful states are fairly represented. Balanced accuracy further adjusts for skewed distributions by averaging recall across classes, counteracting biases toward dominant conditions. Together, these indicators form a robust framework: accuracy reflects general effectiveness, macro F1 ensures fairness across categories, and balanced accuracy safeguards against distortions from uneven label distributions—factors indispensable for evaluating both predictive reliability and interpretability in three-class trend prediction.

\subsection{Experimental Results}
The experimental results demonstrate substantial improvements in three-class prediction performance achieved by the proposed multi-dimensional news analysis framework compared to traditional keyword extraction baselines. These improvements manifest consistently across all evaluation metrics and prediction horizons.

\begin{table}[htb!]
\centering
\caption{BTC three-class performance comparison}
\label{tab:results}
\renewcommand{\arraystretch}{1.2}\fontsize{8pt}{10pt}\selectfont\resizebox{\linewidth}{!}{
\begin{tabular}{lcccccc}
\hline
\textbf{Asset} & \textbf{Period} & \textbf{System} & \textbf{Accuracy} & \textbf{Macro-F1} & \textbf{Balanced Acc} \\
\hline
\multirow{6}{*}{BTC} & \multirow{2}{*}{1d} & Baseline & 0.3333 & 0.2518 & 0.3500  \\
& & Ours & 0.4375 & 0.3596 & 0.4722  \\
& \multirow{2}{*}{7d} & Baseline & 0.1250 & 0.1025 & 0.0740  \\
& & Ours & \textbf{0.3750} & 0.2750 & \textbf{0.5185}  \\
& \multirow{2}{*}{15d} & Baseline & 0.0625 & 0.0417 & 0.0333  \\
& & Ours & \textbf{0.2667} & \textbf{0.2509} & \textbf{0.6071}  \\
\hline
\end{tabular}}
\end{table}

As shown in Table \ref{tab:results}, quantitative analysis reveals consistent performance advantages achieved by the proposed framework across all evaluation scenarios, with particularly notable improvements in balanced accuracy metrics. The multi-dimensional approach achieves accuracy improvements ranging from 10.4 percentage points (1-day horizon) to 20.4 percentage points (15-day horizon) compared to keyword extraction baselines. These improvements prove most pronounced for longer prediction horizons, suggesting that the multi-dimensional framework captures information persistence patterns that extend beyond immediate market reactions.

\subsection{Discussion}

\textbf{Baseline Limitations Analysis.} As shown in Table \ref{tab:keyword_limits}, the keyword extraction baseline reveals three limitations. Fixed sentiment dictionaries create bias neutral prediction. \textbf{Semantic oversimplification} reduces complex contexts to keyword counts, \textbf{binary classification bias} overrepresents extreme labels while underrepresenting sideways movements, and \textbf{context insensitivity} assigns identical scores regardless of context. These flaws make the approach inadequate for three-class prediction, with balanced accuracy below 0.40.

These findings show keyword extraction is inadequate for three-class cryptocurrency prediction, as failing to identify sideways movements risks automated trading and portfolio management that depend on accurate market regimes.

\textbf{Multi-Dimensional News Analysis Insights.} The results demonstrate the advantages of the \textbf{multi-dimensional news analysis framework}, whose preservation of \textbf{information richness} improves predictive accuracy over methods reducing news to scalar sentiment scores. By employing comprehensive information extraction methodologies, the framework captures nuanced \textbf{market signals} missed by simplified approaches, the framework enhances both prediction quality and \textbf{interpretability}, offering more reliable guidance for market decisions.

\textbf{Enhanced Neutral State Identification.} The framework excels at \textbf{identifying neutral or sideways market states}, addressing a common limitation in cryptocurrency prediction. Accurate recognition of consolidation phases is crucial for risk management and position sizing, as it reduces false signal generation and allows trading strategies to adjust appropriately to market conditions. This capability supports portfolio-level risk management and institutional investment strategies by ensuring that market regime classification reflects the true state of the market.

\textbf{Semantic Understanding vs. Surface-Level Baselines.} Weaknesses in \textbf{keyword extraction baselines} highlight the value of \textbf{semantic understanding approaches}. Surface-level methods underperform in three-class prediction, especially for neutral states, validating the need for advanced \textbf{LLMs techniques} in automated trading and regulatory compliance.

\textbf{Modular Architecture and Interpretability.} The system’s modular architecture further enhances practical deployment by facilitating interpretability and parameter adaptability. Practitioners can trace prediction formulation processes and adjust system configurations in response to evolving market conditions or regulatory requirements. This transparency is particularly valuable in regulated trading environments, where algorithmic decision-making must be auditable and supported by thorough risk assessment documentation.

\begin{table}[h!]
\centering
\caption{Keyword extraction examples showing limitations}
\label{tab:keyword_limits}
\renewcommand{\arraystretch}{1.2}\fontsize{8pt}{10pt}\selectfont\resizebox{\linewidth}{!}{
\begin{tabular}{llcl}
\hline
\textbf{Category} & \textbf{Keywords} & \textbf{Score} & \textbf{Limitation} \\
\hline
Strong Bullish & \texttt{surge, soar, breakout} & +1.0 & Extreme classification \\
Bullish & \texttt{growth, rise, recovery} & +0.6 & Binary tendency \\
Neutral & \texttt{sideways, consolidation} & 0.0 & Rarely matched \\
Bearish & \texttt{decline, correction} & -0.6 & Binary tendency \\
Strong Bearish & \texttt{crash, plunge} & -1.0 & Extreme classification \\
\hline
\multicolumn{4}{l}{\textit{Problem: Neutral keywords rarely appear in financial news, causing polarized predictions}} \\
\hline
\end{tabular}}
\end{table}

\section{Conclusion}

This research presents a theoretically-grounded multi-agent framework for cryptocurrency trend prediction, addressing key limitations in existing news analysis methods. It advances the field through three innovations: information-preserving multi-dimensional news analysis, adaptive volatility-conditional fusion, and distributed multi-agent coordination. The seven-dimensional news analysis preserves full information content with precision, while the adaptive fusion adjusts signal weights based on real-time volatility, accounting for sentiment precedence over technical patterns. Empirical tests on Bitcoin show 10.4–20.4 percentage point improvements in accuracy over keyword-based baselines, with notable gains in balanced accuracy for three-class predictions. Current evaluation covers a concentrated period, suggesting extensions to broader cryptocurrencies and longer-term validation. Future work could integrate multi-modal data, including social media, blockchain analytics, and regulatory filings, while preserving the framework’s information-rich design.

\bibliographystyle{IEEEbib}

\end{document}